\documentclass[a4paper,11pt]{article}
\usepackage{graphicx}
\usepackage[colorlinks=true,urlcolor=blue,linkcolor=blue,citecolor=blue]{hyperref}
\usepackage{lineno}
\begin{document}
\title{LA-CoNGA physics:  \\an Open Science Collaboration in Advanced Physics between Latin-America and Europe}
\author{
{\bf Jes{\'u}s Pe{\~n}a-Rodr\'{\i}guez} \thanks{{\tt jesus.pena@correo.uis.edu.co}} \\
{\it Escuela de F\'{\i}sica, Universidad Industrial de Santander,}\\ 
{\it Bucaramanga, Colombia;} \\
{\bf Luis A. N{\'u}{\~n}ez} \\
{\it Escuela de F\'{\i}sica, Universidad Industrial de Santander,}\\ 
{\it Bucaramanga, Colombia.} \\ and
{\it Departamento de F\'{\i}sica, Universidad de Los Andes, }\\
{  M\'{e}rida, Venezuela}\\ \\ on behalf of {\bf LA-CoNGA physics}
}
\maketitle

\abstract{Astrophysics and high energy physicist communities have been pioneers in establishing Virtual Research and Learning Networks, giving rise to productive international consortiums in collaborating environments and training new generations of scientists. 

LA-CoNGA physics (for {\bf L}atin-{\bf A}merican alliance for {\bf C}apacity buildi{\bf NG} in {\bf Advance physics}) is an ERASMUS+ project aiming to support the modernization of university infrastructure and its pedagogical offer in advanced physics in four Latin-American countries: Colombia, Ecuador, Per\'u and Venezuela. This project is co-funded by the Education, Audiovisual and Culture Executive Agency of the European Commission. This virtual teaching and research network comprises three partner universities in Europe and eight in Latin-America; high-level scientific partners (CEA, CERN, CNRS, DESY, ICTP), and several and two industrial partners. 

Open Science education and Open Data are the heart of our operations. In practice, LA-CoNGA physics has created a set of graduate courses in Advanced Physics (high energy physics and complex systems) that are common to all institutions, supported by the installation of interconnected remote laboratories and on an open e-learning platform. This program, incorporated into the master's programs of the eight Latinamerican partners,  is based on three pillars: high energy physics/Complex System courses, data science, and instrumentation.

During 2019 we prepared the syllabuses and selected the lecturers. In 2020 the strict lock-downs modified our pedagogical strategies. The planned model --an eight-node network of universities made-up by local groups for discussions-- was transformed into low-quality home participation. We simplified the connectivity requirements to the minimum bandwidth that could operate remote labs. We also changed the lecture interaction and evaluation model, balancing the teamwork on course projects and continuous evaluation based on class exercises. Despite the lockdown scenario, we managed to support the needs of our instrumentation and computing courses thanks to the contribution and enthusiasm of our partners. With the support of 30 instructors, we gave 100 lectures to 67 students in the four countries. We are now promoting the second cohort due to start in January 2022.
}

{\bf Talk presented at The European Physical Society Conference on High Energy Physics (EPS-HEP2021),
26-30 July 2021. Online conference, jointly organized by Universitat Hamburg and the research center DESY.}
\maketitle

\section{Virtual research and learning community}
A virtual research community is a distributed group of researchers working together with large shared instruments through a virtual environment. These scientific environments --based on high-speed networks, distributed computing, video-conferencing and remote instrument operations-- are commonly referred to as e-Infrastructures. The European Commission has invested heavily in promoting this type of collaboration among scientists worldwide, funding many international projects to create shared e-Infrastructures around the globe~\cite{BordaEtal2006, AndronicoEtal2011}. 

The Latin-American communities of Astrophysics and high energy physicist were pioneers in establishing Virtual Research and Learning Networks, creating productive international consortia in virtual research environments and training the new generation of scientists~\cite{CaicedoEtal2017}. The emerging generation of scientists is trained in this collaborative and distributed open science environment under the virtual research and learning community (VRLC) paradigm~\cite{ArcilacalderonEtal2014}. Employing video-conferencing tools, VRLC develops face-to-face relationships to ensure teacher immediacy and student social presence to facilitate collaborative engagement. Virtual research and learning networks offer accessibility when institutions/groups might not have all the resources/staff locally. Sharing distributed resources provide a collaborative environment that frees educational resources, connectivity, acquisition of digital skills, and foster implementations/development of new learning methods/strategies.  

The COVID-19 pandemic has catalysed many of these practices in the academy. We learned that we could manage much of these teaching activities. We became convinced that ideas can flow without being driven by mobility and physical attendance at congresses and/or scientific meetings. We gradually realised that many of these learnings are here to stay. When the pandemic passes, the geography and, above all, the dynamics of research groups will be different. We will have other practices and, in many cases, we will be different and more effective.

This short paper will discuss the concept and early impacts of one particular implementation of a VRLC in four Andeans countries: LA-CoNGA Physics ({\bf L}atin-{\bf A}merican alliance for {\bf C}apacity buildi{\bf NG} in {\bf Advance physics}~\footnote{\url{https://LA-CoNGA.redclara.net/}}). In the next section, we briefly describe the antecedents. In section \ref{LACoNGAPhysics} we detailed the project: its syllabus (section \ref{Syllabus}), the learning environment (section \ref{Environment}), the remote labs infrastructure (section \ref{RemoteLABs}) and the sustainability strategies (section \ref{Sustainability}). We end with few final remarks in section \ref{Conclusions}.

\section{High Energy, Astrophysics and Astroparticle Physics in Latin-America}

The high-performance computing, High Energy, Astrophysics, Cosmology and Astroparticle Physics communities have grown in Latin-America in the last decades. Several international scientific alliances have been established throughout the continent, and the participating institutions provide technology development, academic formation, financial investment, and modernisation of STEM education at the region~\cite{GitlerGomesNesmachnow2020, Dib2021}. A recent initiative, the Latin-American Strategy Forum for Research Infrastructure for High Energy, Cosmology and Astroparticle Physics (LASF4RI-HECAP \footnote{\url{https://lasf4ri.org/}}) submitted a declaration to the IV Iberoamerican Science and Technology Ministerial Meeting. 
This document presents a plan of action including recommendations and a roadmap for these disciplines~\cite{AiharaEtal2021}.

Three key training actions can be considered a precedent for the LA-CoNGA Physics master program: the High Energy Physics Latinamerican European Network (HELEN)~\footnote{\url{https://www.roma1.infn.it/exp/helen/}}, the European Particle physics Latin-America NETwork (EPLANET) and Centro Virtual de Altos Estudios en Altas Energías (CEVALE2VE)~\footnote{\url{http://www.cevale2ve.org/en/home/}}. HELEN creates an educational-scientific network among universities and institutes in Latin-America and Europe to train the new generations of physicists in High Energy Physics~\cite{Maiani2005}. Next  EPLANET develops the technical skills of young scientific personnel by their participation in top experiments at CERN and the Pierre Auger Observatory~\cite{Maiani2014}. Finally, we introduced the CEVALE2VE initiative, an online HEP community oriented to modernise the academic environment in Latin-American, by stimulating physics' students in the scientific careers~\cite{CaicedoEtal2017}.

The development of the above mentioned areas in Latin America is nuanced and variable country-by-country. However, it has enormous potential thanks to the diversity of interests/skills and a young generation with potential and eagerness to learn.

\section{LA-CoNGA Physics}
\label{LACoNGAPhysics}
\begin{figure}[h!]
    \centering
    \includegraphics[width=0.35\textwidth]{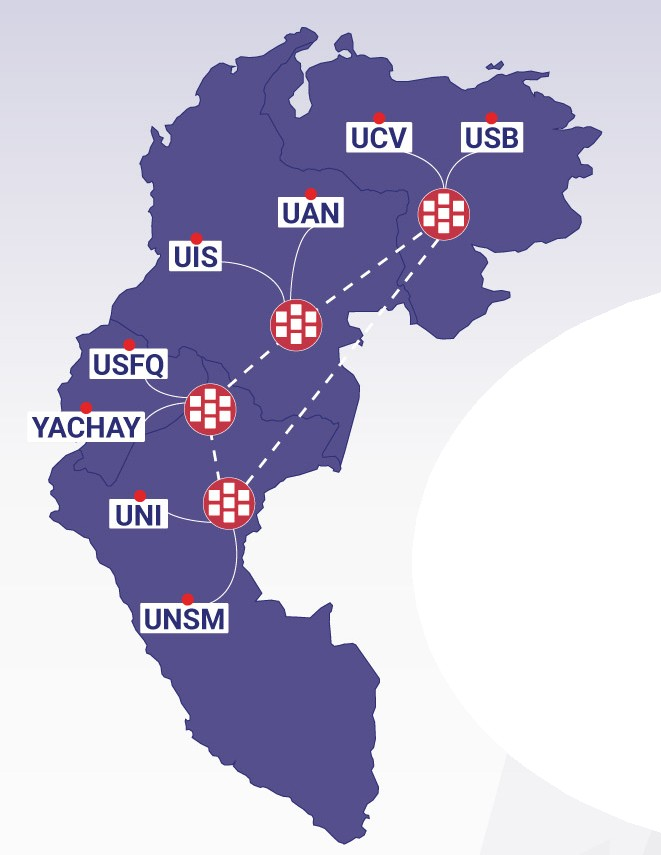}
    \caption{LA-CoNGA physics aims to update teaching environments in eight Latin-American HEIs from Colombia, Ecuador, Perú and Venezuela.    Latin-American universities participating in LA-CoNGA physics: Universidad Central de Venezuela (UCV) and Universidad Simón Bolívar (USB) from Venezuela; Universidad Antonio Nariño (UAN) and Universidad Industrial de Santander (UIS) from Colombia; Universidad San Francisco de Quito (USFQ) and Universidad Yachay (YACHAY) from Ecuador; and Universidad Nacional de Ingeniería (UNI) and Universidad Nacional de San Martín (UNSM) from Perú.}
    \label{fig:LACoNGAUniversities}
\end{figure}

The main objective of LA-CoNGA physics is to update teaching environments in eight Latin-American higher education institutions (HEI) from Colombia, Ecuador, Perú and Venezuela (see Figure \ref{fig:LACoNGAUniversities}). Using Advanced Physics as a model, the proposed modernisation implements an innovative syllabus and learning atmosphere using standard research conditions and tools. Implementing the Bologna model, we offer a one-year Master's program, oriented to problem-solving. With a planned sustainability strategy in mind, the courses are inserted within the local graduate studies of each target institution, with the intention of strengthening the cross-institutional relations among the target HEIs.

\subsection{A combined syllabus}
\label{Syllabus}
The LA-CoNGA physics syllabus is a combination of three fundamental modules: Theory, Data Science and Scientific Instrumentation. The Theory module presents a framework for Advanced Physics concepts applied to High Energy Physics and Complex Systems~\footnote{\url{https://LA-CoNGA.redclara.net/courses/}}. Simultaneously, the Data Science and Scientific Instrumentation modules provide basic techniques and skills used in (and outside) the physics career. Two optional courses and a three-month internship complement the introductory courses, expanding the vision by applying the concepts and techniques discussed in the three modules.

\begin{figure}[h!]
    \centering
    \includegraphics[width=0.65\textwidth]{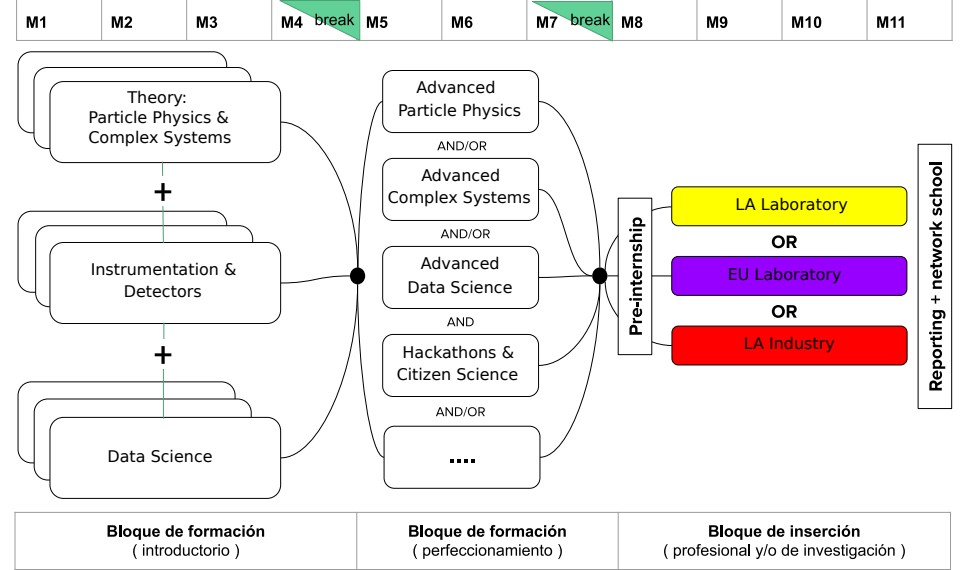}
    \caption{LA-CoNGA physics syllabus consists of three fundamental components: Theory, Data Science and Scientific Instrumentation. The student's academic training begins in the first two periods, and the professional/research insertion comes in the last one}
    \label{fig:imple}
\end{figure}

LA-CoNGA implements its program in three periods per year as shown in Figure \ref{fig:imple}. We developed a three-track skill training (on equal footing) in the first period: Theory, Scientific Instrumentation and Data Science. 

The first semester is mainly composed of mandatory courses and hands-on activities, with each track having 10 ECTS~\footnote{The European Credit Transfer and Accumulation System (ECTS) make it easier for European students to move between countries having their academic qualifications and study periods recognised.} assigned. The second semester is split into four 8-week-long courses,  followed by a 14-week-long research internship. The courses are composed as follows: two are mandatory and common to both HEP and CS (``Advanced topics in data science'' and ``Introduction to Medical Physics''), and one specific to each track (``Astro-particles and Cosmology'' for HEP and ``Advanced statistical mechanics'' for CS). Each course has 5 ECTS assigned (yielding, therefore, a total of 15 ECTS per student). The research internship also has 15 ECTS assigned and is structured as follows: an initial pre-internship report to be completed after two weeks (usually a bibliographical survey), the internship work, and finally it's written report with an oral presentation during the School Network at the end of the school year.

A two-prong training in advanced physics is offered: either on High-Energy physics (HEP) or Complex Systems (CS). The two tracks are reflected in each of the three skill alternative via specific courses and hands-on activities. The main course in the Theory block: ``Introduction to Field Theory'', follows an innovative approach, with a first 8-week long subcourse common to all students, and later split into two different HEP and CS subcourses (also 8-week long). The initial subcourse is built to present a shared set of concepts and tools to HEP and CS students. It provides an accepted way for a modern understanding of many cutting-edge phenomena, i.e. spontaneous symmetry breaking and renormalisation group.

Twice a month~\footnote{\url{https://www.youtube.com/hashtag/seminariosLA-CoNGA}}, the LA-CoNGA community follow open-access virtual seminars, where experts working in industry and academy share their experiences with students and colleagues.  In the first period, 67 students from the four countries followed more than 100 hours of courses, given by 30 instructors from Latin-America and Europe. We implemented surveys to evaluate the courses and remote-lab quality to get feedback to improve LA-CoNGA content and platform as an iterative process. LA-CoNGA remote-labs were accessed for more than 100 hours by students carrying out their HEP final class projects.

The second period has courses in Statistical Mechanics, Reproducibility in Science, Astroparticles \&
Cosmology, and Medical Physics.

\subsection{The learning environment}
\label{Environment}
The learning environment relies on a e-platform in-house developed based on standard open-access tools commonly available for research collaborations in High Energy Physics and Astrophysics. These tools favour the replicability of the scientific process, by sharing the data and codes associated with creating knowledge. Finally, remote instrumentation and interconnected laboratories complete this real-life operational research environment.

The LA-CoNGA learning suite integrates a Git repository for version control of source codes and documents, a data repository to preserve and give access to samples from instrumental measurements and computer-generated data, and also includes an instant messaging tool and a computational interface. The instant messaging system organizes and stores team discussions. The computational interface allows the users to process and analyze data from the repository or directly from an instrument. All codes and generated data, saved into git and/or data repositories, can be shared among users.

The pandemia changed all our pedagogical strategies. The planned model was an eight-node network of universities assuming local groups for discussions and interactions. The instructor would help support the course evaluation of local students and cooperate with the evaluation of the other institutions. The university e-infrastructure guarantees good connectivity among the nodes and supports the remote operation of lab equipment. 

The house lockdown and low-quality home connections modified all the interactions among  students, classes and the lab infrastructure. We simplified the connectivity requirements to a minimum bandwidth that could cope with the remote equipments. We have also been forced to change the model of interaction and evaluation, balancing the teamwork on course projects and continuous evaluation based on class exercises. 

\subsection{A network of remote laboratories and computational facilities}
\label{RemoteLABs}
In addition to the global sanitary situation, we have faced severe administrative difficulties with the purchase capability of the consortium: most of the equipment needed for our project are instruments for Nuclear and Particle Physics produced by few manufacturers worldwide.  

Despite this unfavourable scenario, we managed to support the needs of our instrumentation and computing courses based on the contribution and enthusiasm of our partners.  We implemented remote laboratory setups based on detectors available in Bucaramanga, Colombia.  From UIS, we allowed the students to perform online remote measurements on two particular experiences. A local technical staff configured and maintained the functioning detectors, and the connected students could remotely operate the sensors and collect the data. 

CEDIA (for Corporación {\it Ecuatoriana para el Desarrollo de la Investigación y la Academia}, the Ecuador Academic Network~\footnote{\url{https://www.cedia.edu.ec}}), in cooperation with CLARA (for {\it Cooperación Latinoamericana de Redes Avanzadas}~\footnote{CLARA is a multilateral organization strengthen the development of science, education, culture and innovation in Latin America, through the innovative use of networks, infrastructure and advanced information technologies \url{https://redclara.net/}}), provided the computational resources covering needed by our students for the Data Science and Theory modules.  These resources were used for coding and computing, guaranteeing the day-to-day interaction with the academic community and course content management. Finally, UIS also provided video conferencing resources for all the online classes.

The original plan was to train students connected locally via the network resources available at their university campuses.  But because all of them and most of our instructors had to operate from their homes during the lockdown periods. The information technology infrastructure was often minimal from home sites, making it difficult for our students to interact among them and access the experimental setups. The other unfortunate situation was the impossibility to access the institutional laboratories and classrooms. We partially solved this difficulty by creating a library of video resources complementing the recorded classes. The student with connectivity problems could follow the course asynchronously when their connection was reestablished.

\begin{figure}[h!]
    \centering
    \includegraphics[width=0.45\textwidth]{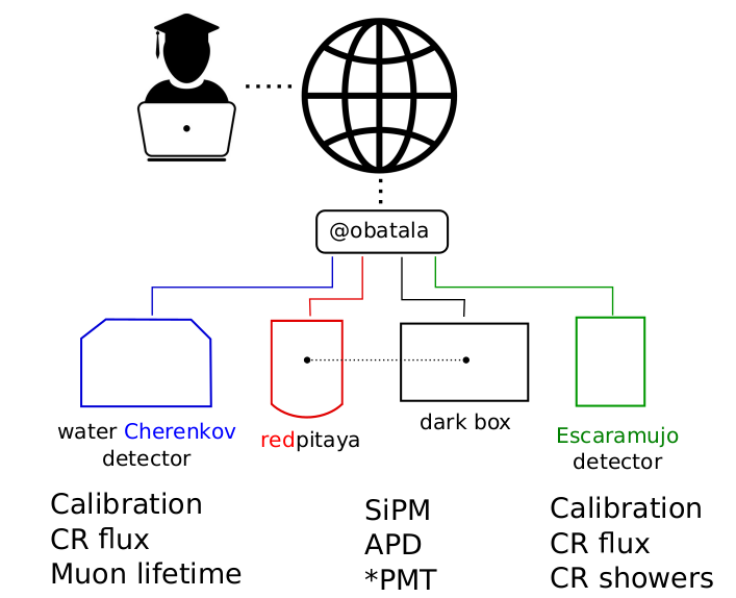}
    \caption{LA-CoNGA implements a proof of concept of remote-labs to develop physics projects operating under the low bandwidth home connections. Students access the remote-lab equipment by SSH from home. They can carry out calibration, data taking, and DAQ parameter control tasks.}
    \label{fig:remlab}
\end{figure}

Management tasks from the lab side are needed to guarantee an excellent remote lab. Such activities include scheduling the experiments, user authentication, user interfaces and statistics (number of users, location, connection time, results and reports).

Our remote labs follow a series of additional steps to fulfil the requirements of face-to-face lab work. To generate an alternative due to the lockdown interaction, we developed a protocol. First, the students receive the bibliography on the experiment and with a video, we introduce the idea behind the actual experimental setup. Secondly, the students access the remote lab from home for taking data by controlling acquisition parameters (gain, voltages, temperature, among others). Finally, they analyse the collected data and report the results to their classmates.

LA-CoNGA remote lab facilities are accessed by using SSH protocol as shown in Figure \ref{fig:remlab}. The remote lab has equipment for evaluating SiPM/APD, scintillator and water Cherenkov detectors. Students can work with the equipment emulating the different development stages of a high energy physics experiment: calibration, measuring and data analysis.

\subsection{Sustainability strategy}
\label{Sustainability}
The sustainability strategy aims to establish academic interrelations among the consortium stakeholders outside the scholarly activities related to the LA-CoNGA physics syllabus.

The project promotes joint seminars among the different Master's programs and cooperation among the international relations offices to generate common quality assurance indicators and share the educational resources provided by the project.

The courses have been included in six of the eight Master's programs. For some universities, this involved a significant change in their existing programs, updating and reforming their syllabus to include the new areas like Data Science and Instrumentation offered by LA-CoNGA physics. These courses have profited from more than two dozen lecturers, which have provided the students with a broad spectrum of visions and projects.

All consortium members play a crucial role in implementing the project, with well-defined leadership responsibilities and tasks. This ensures they acquire/strengthen the needed skills, know-how the project works internally and generate internal connections. 

A code of conduct, a diversity policy, and an open data plan are three crucial ingredients of the region's LA-CoNGA Physics vision for capacity building. The consortium implements these three actions and discusses them with the students and the international relations offices of the partner HEIs.

Finally, several LA-CoNGA physics members have actively participated in the LASF4RI-HECAP process contacting policymakers and funding agencies to create a roadmap for developing High Energy, Cosmology and Astroparticle Physics for the first time in Latin America. This membership opens the opportunity to generate networking that may find funding opportunities to continue this project incorporating new institutions and extending the scope to other disciplines.

\section{Conclusions}
\label{Conclusions}
We created a sustainable, dynamic, collaborative, interconnected, and diverse virtual research and learning network of advanced physics Latin-American and European students/researchers.

The learning environment developed in-house for the consortium, made of professional open-source tools, trains our students to work in actual research conditions and enables the consortium to maintain daily interactions among our members.  Now, this collaborative platform is considered a candidate to be a stand-alone platform to help small and medium research groups operate.

We have produced a proof of concept by installing and operating remote equipment to perform physics lab projects under the low bandwidth required by home connections. These implementations are now fully operational as a contribution of one of the partner (UIS) institutions. 

LA-CoNGA contributes to the modernisation, accessibility, and internationalisation of higher education systems in the region by using technology and research practices in educational environments, enhancing learning processes, applying good scientific practices and gender equality. We envision similar experiences in other disciplines across our continent.

%\bibliographystyle{unsrt}
%\bibliography{references.bib}

\end{document}